\documentclass[12pt]{article}
\newcommand{\be}{\begin{equation}}
\newcommand{\ee}{\end{equation}}
\newcommand{\bea}{\begin{eqnarray}}
\newcommand{\eea}{\end{eqnarray}}
\newcommand{\nn}{\nonumber \\}
\newcommand{\cN}{{\cal N}}
\newcommand{\p}[1]{(\ref{#1})}
\def\({\left(}
\def\){\right)}
\def\a{\alpha}
\def\b{\beta}

\def\d{\delta}

\def\L{\Lambda}

\def\ve{\varepsilon}

\def\tr{{\rm tr}}

\def\nb{\nabla}

\begin{document}

\thispagestyle{empty}

\begin{center}
{\Large \bf Hidden supersymmetry as a key to constructing \\
\vspace{0.2cm}

low-energy superfield effective actions}
\end{center}

\begin{center}

{\bf I.L. Buchbinder${}^{a}$\footnote{joseph@tspu.edu.ru},  E.A.
Ivanov${}^b$\footnote{eivanov@theor.jinr.ru}
} \\
\vspace{5mm}

\footnotesize{ ${}^{a}${\it Department of Theoretical Physics,
Tomsk State Pedagogical University, 634061, Tomsk, Russia} \\
${}^{b}$ {\it Bogoliubov Laboratory of Theoretical Physics, JINR,
141980 Dubna, Moscow region, Russia}}
\vspace{2mm}

\end{center}

\begin{abstract}
In this review paper, we outline and exemplify the general method of constructing the
supefield low-energy quantum effective action of supersymmetric
Yang-Mills (SYM) theories with extended supersymmetry in the Coulomb
phase, grounded upon the requirement of invariance under the non-manifest (hidden)
part of the underlying supersymmetry. In this way we restore
the  $\cN=4$ supersymmetric effective actions in $4D, \cN=4$ SYM,
$\cN=2$ supersymmetric effective actions in $5D, \cN=2$ SYM and
$\cN=(1,1)$ supersymmetric effective actions in $6D, \cN=(1,1)$ SYM
theories. The manifest off-shell fractions of the full supersymmetry
are, respectively, $4D, \cN=2$, $5D, \cN=1$ and $6D, \cN=(1,0)$
supersymmetries. In all cases the effective actions depend on the
corresponding covariant superfield SYM strengths  and the
hypermultiplet superfields. The whole construction essentially
exploits a power of the harmonic superspace formalism.
\end{abstract}
\vspace{1.5cm}

\begin{center}
{\it Contribution to the Volume in Honor of Andrei A. Slavnov \\ on the
occasion of his 80th Birthday}
\end{center}

\vfill
\newpage

\setcounter{page}{1}
\setcounter{footnote}{0}

\section{Introduction}
Supersymmetry is still a source of the various surprises in
theoretical and mathematical physics. It is sufficient to say that
supersymmetry allows to construct the completely finite models of
quantum field theory, to formulate the phenomenologically attractive
models beyond the standard model, to eliminate ghosts in the
spectrum of string theory, to obtain the exact results in quantum
mechanics and in the classical and quantum field theory, etc. Among
a lot of works on supersymmetry, we wish to distinguish the papers
by A.A. Slavnov \cite{Slav1,Slav2,Sl2,Sl3}, which are directly
or indirectly related to the problem of effective action.

Quantum effective action is a central object of quantum field
theory, and it is used for the study of plenty aspects of the latter, such as
renormalization, calculation of $S$-matrix amplitudes, finding the
quantum corrections to the classical equations of motion, dynamical
symmetry breaking,  symmetries of quantum non-abelian gauge theories
(they are described by Slavnov-Taylor identities \cite{Tay,S}) and
many others ( see e.g., \cite{QFT2}).

The low-energy effective action plays an important role in
supersymmetric gauge theories, providing a link between
superstring/brane theory and quantum field theory. On the one hand,
such an effective action can be calculated in the quantum field
theory setting and, on the other, it can be derived within the brane
stuff. As a result, the low-energy effective action allows one, in
principle, to describe the low-energy string effects by methods of
quantum field theory and vice versa (see reviews
\cite{BUCH1,BUCH2,BUCH3}).

The most elegant way to study the quantum structure of the
supersymmetric field theories is through their  formulations in terms
of unconstrained superfields, which secure a
manifest supersymmetry at all stages of calculations. Such a formulation is well
developed for $4D, \cN=1$ supersymmetry (see, e.g.,
\cite{B&K}). However, for higher-dimensional and
extended supersymmetric theories the formulation in terms of
unconstrained superfields faces some problems, and it was worked out only
for a few special cases. One of these cases is $4D, \cN=2$
supersymmetry, where the successful formulation in terms of
unconstrained superfields was realized in terms of harmonic
superspace \cite{GIOS}. Using this approach allows  one to formulate
the four-dimensional maximally extended $\cN=4$ gauge theory in such a
way that two supersymmetries are manifest and two others are
on-shell and hidden (non-manifest). As a result, we arrive at $\cN=4$
supersymmetric formulation of the theory under consideration in
terms of $\cN=2$ harmonic superfields.

Extended supersymmetry imposes severe constraints on the classical
and effective quantum superfield actions of gauge theories.
A good example is the four-derivative term in the low-energy $4D,
\cN=4$ SYM effective action (in the Coulomb phase) which, in the
sector of $\cN=2$ gauge multiplet, is accommodated by a
non-holomorphic superfield potential \cite{DS}. An $\cN=4$
supersymmetric completion of this potential by the hypermultiplet
terms in $\cN=2$ harmonic superspace was constructed for the first
time in our paper \cite{BuIv} on the purely symmetry grounds. It was
further reproduced in \cite{BIP} from the quantum $\cN=2$ supergraph
techniques. The origin of non-renormalizability of the $\cN=4$ SYM low-energy
effective action against higher-loop quantum corrections was
established and links with the leading terms in the effective action
of D3 brane on the $AdS_5\times S^5$ background were indicated (see
also \cite{BBP,BP} and \cite{BUCH3} for a review).

The key observation of \cite{BuIv} consisted in that the constraints of the
second, hidden ${\cal N}=2$ supersymmetry completing the manifest
${\cal N}=2$ supersymmetry to the total (on-shell) ${\cal N}=4$ one
are so strong that fix the form of the relevant
superfield potential up to an overall coefficient, which further has
to be calculated  from quantum considerations (like in \cite{BIP})
\footnote{Earlier, the importance of taking into account the total supersymmetry for studying
the effective action of extended supersymmetric gauge theories was
pointed out in \cite{BKT}.}.

Later on, a similar approach was applied in $3D$ gauge theories,
where it allowed one to determine the leading quantum corrections in
$\cN=4$ SYM theory \cite{BPS09,BPS10}, and to construct $\cN=3$
superfield ABJM action \cite{N3ABJM}. It also turned out  useful for
revealing the structure of the leading terms of the effective action
in $2D$ gauge theories with extended supersymmetry \cite{S17}. As
the latest application of the method invented in \cite{BuIv}, the
complete structure of the leading terms in the low-energy effective
actions of $5D, {\cal N}=2$ and $6D, {\cal N}=(1,1)$ SYM theories
was established  \cite{5DBIS,Buchbinder:2017xjb}.

In this review paper we explain the basics of our method from the
single point of view, starting from the original example of ref.
\cite{BuIv} and then proceeding to the recent results of refs.
\cite{Buchbinder:2017xjb, 5DBIS}.  We point out the decisive role of
the harmonic superspace (HSS) approach
(\cite{GIKOS,GIOS1,GIOS2,GIOS}) for the derivation of the complete
superfield effective actions in the situations, when the superfield
description with {\it all} relevant supersymmetries manifest and
off-shell is still unknown.

The paper is organized as follows. In \textbf{section 2} we discuss
the $\cN=4$ supersymmetric low-energy effective action in $4D,
\cN=4$ SYM theory. This theory is formulated in harmonic
superspace in terms of the gauge multiplet and hypermultiplet
superfields. The theory exhibits the manifest
off-shell $\cN=2$ supersymmetry, as well as the second additional $\cN=2$ supersymmetry, which is
non-manifest (hidden) and forms, together with the manifest  $\cN=2$ supersymmetry, the whole $\cN=4$
supersymmetry only on shell. Then we construct a quantum low-energy effective action
in such a theory in the Coulomb phase. We start from the effective
potential in the gauge multiplet sector calculated earlier in a series
of papers \cite{PvU,G-rR,BBKO,BK,5,GKPR,BBK,LvU,BKT,KM,Kuz01,b,G-R},
and show how this result can be completed by the hypermultiplet terms
to the effective potential depending on all fields of $\cN=4$ gauge
multiplet. This completion is derived algebraically, solely  on the basis of
the extra on-shell $\cN=2$ supersymmetry, and so demonstrates the power of hidden
supersymmetry for such calculations.

\textbf{Section 3} is devoted to the derivation of the low-energy
effective action in $5D, \cN=2$ supersymmetric gauge theory. Like in
section 2, we begin with the classical formulation of the theory
in harmonic superspace, where half of the
supersymmetries is realized manifestly and another half in a hidden
way. Then we study the structure of the low-energy effective action
using both manifest and hidden supersymmetries. Here it is worth
discussing an important point. The $5D, \cN=1$
gauge theory admits a classical Chern-Simons action, however its
$\cN=2$ supersymmetric generalization does not exist since in
$\cN=1$ case the Chern-Simons action respects the invariance under
$5D, \cN=1$ superconformal algebra $F(4)$ which is unique and
possesses no higher $\cN$ extension. It is possible to show, by direct quantum computations in $5D,
\cN=1$ superspace \cite{Pletnev}, that the two-derivative
contributions (the Chern-Simons term) to the $\cN=2$ SYM effective
action coming from the hypermultiplet and from the ghost superfields
precisely cancel each other. This cancelation is analogous to the
well-known phenomenon in $3D$ case \cite{BPS09,BPS10}, where
Chern-Simons term cannot arise as a quantum correction to the
effective action in supersymmetric gauge theories with $\cN>2$.

As we demonstrate in this section, the four-derivative term, on the contrary, admits the unique hypermultiplet
completion under the requirement of an implicit $5D, \cN=1$ on-shell supersymmetry alongside with the
manifest off-shell $\cN=1$ one. The procedure of constructing such a
hypermultiplet completion is quite analogous to the one in
\cite{BuIv}.

In \textbf{section 4} we address the problem of construction of the low-energy
effective action in $6D, \cN=(1,1)$ SYM theory. Such a theory is
formulated in $6D, \cN=(1,0)$ harmonic superspace as the theory of
interacting $\cN=(1,0)$ gauge multiplet and hypermultiplet, both in
the adjoint representation of gauge group. The theory possesses the manifest $\cN=(1,0)$ supersymmetry and
an additional hidden $\cN=(0,1)$. On shell they close on the full
$\cN=(1,1)$ supersymmetry. Exploiting the hidden supersymmetry, we find
the effective action in the gauge multiplet sector basically on the symmetry
grounds.

In \textbf{section 5} we list the basic results and discuss some
further possible developments.

\setcounter{equation}{0}
\section{Low-energy effective action of $4D\,, \cN=4$ SYM theory}
It is known that D3-branes are related to $4D, \cN=4$ SYM theory
(see, e.g., \cite{GK,BKLS}). Interaction of D3-branes is described
in abelian bosonic sector by the Born-Infeld action, the
leading low-energy correction being of the form $\sim \frac{F^4}{X^4}$,
where $F^4$ denotes a structure of fourth degree in an abelian field
strength $F_{mn}$ and $X$ stands for the scalar fields of $4D,\,
\cN=4$ gauge (vector) multiplet. The one-loop calculation of such an
effective action in the Coulomb branch of $\cN=4$ SYM theory, both
in the component approach and in terms of  $\cN=1,2$ superfields,
has been accomplished in ref.
\cite{PvU,G-rR,BBKO,BK,5,GKPR,BBK,LvU,BKT,KM,Kuz01}. The complete
$\cN=4$ structure of the one-loop low-energy effective action has
been found in \cite{BuIv,BIP}. The two-loop contributions to
the low-energy effective actions of $\cN=4$ SYM theory have been
considered in \cite{BPT,KUA,KU}. The structure of the low-energy
effective action in the mixed Coulomb - Higgs branch was a subject
of ref. \cite{BP07}. A review of the results related to the
calculations of low-energy effective actions in $4D$
extended supersymmetric gauge theories can be found, e.g.,  in
\cite{BUCH1,BUCH2} \footnote{Various aspects of $\cN=2$
harmonic superfield models were also discussed in \cite{BKO}-\cite{BS1}.}.

Studying the low-energy effective action \footnote{By the
low-energy effective action we always mean the leading in the external
momenta piece of the full quantum effective action.}
of ${\cal N}=4$ SYM models was initiated in \cite{DS}. In the
${\cal N}=2$ superfield formulation, the full ${\cal N}=4$ gauge multiplet is constituted
by the ${\cal N}=2$ gauge multiplet and hypermultiplet.
The authors of \cite{DS} studied
the effective action of ${\cal N}=4$ SYM theory with the gauge group SU(2)
spontaneously broken to U(1) and considered that part of this action which
depends only on the fields of massless U(1) ${\cal N}=2$ gauge multiplet.
The requirements of scale and R-invariances specify this part of the
effective action up to a numerical coefficient. The result can be given in
terms of non-holomorphic effective potential
\be
{\cal H}(W,\bar{W})=c\,\ln\frac{W}{\Lambda}\,\ln\frac{\bar{W}}{\Lambda}~,
\label{1}
\ee
where $W$ and $\bar{W}$ are the ${\cal N}=2$ superfield strengths,
$\Lambda$ is an
arbitrary scale and $c$ is an arbitrary real constant. The effective action
defined
as an integral of ${\cal H}(W,\bar{W})$ over the  full  ${\cal N}=2$ superspace
with the coordinates $z = (x^m, \theta_{\alpha i},
\bar\theta^i_{\dot\alpha})$ is independent of the scale $\L$. It is
worth to point out that the result (\ref{1}) was obtained in ${\cal N}=4$ SYM theory
entirely on the symmetry grounds \footnote{Non-holomorphic potentials of the
form \p{1} as possible contributions to the effective action in ${\cal N}=2$ SYM
theories were earlier considered in refs. \cite{b}.}.

Eq. (\ref{1}) provides the {\it exact} form of the low-energy effective
action of ${\cal N}=4$ SYM theory in the ${\cal N}=2$ gauge superfield sector. Any quantum corrections can be absorbed into
the coefficient $c$. One can show \cite{DS,LvU}
that the non-holomorphic effective potential (\ref{1}) receives neither
perturbative nor
non-perturbative contributions beyond one loop. As the result,
construction of exact
low-energy effective action for SU(2) SYM theory in the Coulomb
branch [i.e. with SU(2) broken down to U(1)] is reduced to computing the
coefficient $c$  in the one-loop approximation.

The direct derivation of the potential (\ref{1}), computation of the
coefficient $c$
and, hence, the final reconstruction of the full exact low-energy U(1)
effective action in the gauge field sector
from the quantum ${\cal N}=4$  SYM theory were undertaken in refs.
\cite{PvU,G-rR,BK}. In particular, it was found that $c=\frac{1}{(4\pi)^2}$.
Further studies showed that the result (\ref{1}) for the gauge
group SU(2) spontaneously broken to U(1), can be generalized to the group
SU(N) broken to its
maximal abelian subgroup \cite{5}-\cite{LvU}. The relevant one-loop effective
potential is given by
\be
{\cal H}(W,\bar{W})=c\sum\limits_{I<J}\ln\frac{W^I-W^J}{\Lambda}\,
\ln\frac{\bar{W}^I-\bar{W}^J}{\Lambda}~,  \label{2}
\ee
with the same coefficient $c$ as in (\ref{1}) for SU(2) group. Here
$I,J=1,2,\dots,N$, $W=\sum\limits_1W^Ie_{II}$ belongs to Cartan
subalgebra of the algebra su(N), $\sum\limits_iW^I=0$, and $e_{IJ}$ is
the Weyl basis in su(N) algebra (for details see ref. \cite{BBK}).

\subsection{Action of ${\cal N}=4$ SYM theory in ${\cal N}=2$ HSS}
The ``microscopic'' action of ${\cal N}=4$ SYM theory in the formulation through ${\cal N}=2$ harmonic superfields can be written as
\be
S[V^{++},q^+]=
\frac{1}{8}\left(\int d^8\zeta_L {\rm tr\, W^2}+
\int d^8\zeta_R {\rm tr\, \bar{W}^2}\right)-
\frac{1}{2}\int d\zeta^{(-4)}{\rm tr\,}q^{+a}
\left(D^{++}+igV^{++}\right)q^+_a~.
\label{4}
\ee
The real analytic superfield $V^{++}$ is the harmonic gauge potential
of ${\cal N}=2$ SYM theory and the analytic superfields $q^+_a, \;a=1,2~,$
describe the hypermultiplets (they satisfy the pseudo-reality condition
$q^{+a} \equiv \tilde{q}^+_a = \ve^{ab}q^+_b$, where the generalized
conjugation $\,\sim\,$ was defined in \cite{GIKOS}).
  The ${\cal N}=2$
superfield strengths $W$ and $\bar{W}$ are expressed in terms of
$V^{++}$. The superfields $V^{++}$ and $q^{+}_a$ belong to adjoint
representation of the gauge group,  $g$ is a coupling constant,
$d^8\zeta_L=d^4xd^2\theta^+d^2\theta^-du$,
$d^8\zeta_R=d^4xd^2\bar{\theta}^+d^2\bar{\theta}^-du$ and
$d\zeta^{(-4)}=d^4xd^2\theta^+d^2\bar{\theta}^-du$ are the measures of integration over chiral, anti-chiral  and
harmonic analytic ${\cal N}=2$ superspaces, $du$ is the measure of integration over the harmonic variables $u^{\pm\,i},
\;\;u^{+ i}u^-_i = 1$. Any further details regarding the action
(\ref{4}), in particular, the precise form of the
analyticity-preserving harmonic derivative $D^{++}$, can be found in
refs \cite{GIKOS} - \cite{GIOS2}, \cite{GIOS}. We shall basically
follow the notation of the book \cite{GIOS}.

Either term in (\ref{4}) is manifestly ${\cal N}=2$ supersymmetric.
Moreover, the action (\ref{4}) possesses an extra hidden ${\cal
N}=2$ supersymmetry which mixes up $W$, $\bar{W}$ with $q^{+}_a$
\cite{GIOS1,GIOS2,GIOS,BBK}. As a result, the model under
consideration is actually ${\cal N}=4$ supersymmetric. Our aim is to
examine the possibility of constructing ${\cal N}=4$ supersymmetric
functionals whose $q^+$-independent parts would have the form of
(\ref{1}), (\ref{2}).

The effective potentials (\ref{1}) and (\ref{2}) involve chiral and
anti-chiral abelian strengths $W$ and $\bar{W}$ satisfying the free
classical equations of motion $(D^+)^2W= (\bar D^+)^2 \bar W = 0$,
where the harmonic projections of the spinor ${\cal N}=2$
derivatives $D^i_\alpha, \bar D^i_{\dot\alpha}$ are defined as
$D^\pm_\alpha = D^i_{\alpha}u^\pm_i,  \bar D^\pm_{\dot\alpha} = \bar
D^i_{\dot\alpha}u^\pm_i.$ So, in order to construct the above
functionals we need to know the hidden ${\cal N}=2$ supersymmetry
transformations only for on-shell $W, \bar W$ and, respectively, for
on-shell $q^{+a}$ ($D^{++}q^{+ a}=0$). For further use, it is
instructive to write down the complete set of equations for the
involved quantities, both on and off shell. \underline{Off-shell}:\,\, $\bar
D^\pm_{\dot\alpha} W =  D^\pm_\alpha \bar W = 0~, \quad (D^\pm)^2W =
(\bar D^\pm)^2 \bar W~,\,\, D^+_\alpha q^{+ a} = \bar
D^+_{\dot\alpha} q^{+ a} = 0~.$ \underline{On-shell}: $(D^\pm)^2 W = (\bar
D^\pm)^2 \bar W = 0~, D^{++} q^{+a} = D^{--} q^{-a} = 0, \, q^{- a}
\equiv D^{--} q^{+ a}, \, D^{++} q^{- a} = q^{+a}, \, D^-_\alpha
q^{- a} = \bar D^-_{\dot\alpha} q^{- a} = 0.$ In checking the
on-shell relations for the hypermultiplet superfield an essential
use of the commutation relation $[D^{++}, D^{--}] = D^0$ should be
made, with $D^0$ being the operator which counts harmonic $U(1)$
charges, $D^0q^{\pm a} = \pm q^{\pm a}$.

It is known that, in the central basis of the harmonic superspace,
\be q^{\pm a} = q^{ia}(z)u^\pm_i~, \label{onshq} \ee where
$q^{ia}(z)$ is the on-shell hypermultiplet superfield independent of
harmonic variables and defined on the standard ${\cal N}=2$
superspace with the coordinates $z = (x^m , \theta_{\alpha i}, \bar
\theta^i_{\dot\alpha})$.  Note that in this on-shell description,
harmonic variables are to some extent redundant, everything can be
formulated in terms of ordinary ${N}=2$ superfields $W(z), \bar W
(z), q^{ia}(z)$. The use of the harmonic superspace language is
still convenient, e.g., because of the possibility to integrate by
parts with respect to the harmonic derivatives in the effective
action.

Taking into account these remarks, the on-shell form of the
hidden ${\cal N}=2$ transformations can be written as  \cite{GIOS} \bea && \delta
W = {1\over 2}\bar\epsilon^{\dot\alpha a}\,\bar
D^-_{\dot\alpha}q^+_a\;, \quad \delta \bar W = {1\over
2}\epsilon^{\alpha a}\, D^-_{\alpha}q^+_a~, \nn && \delta q^+_a
={1\over 4}\,(\epsilon^\beta_a D^+_\beta W +
\bar\epsilon^{\dot\alpha}_a\bar D^+_{\dot\alpha} \bar W)~, \quad
\delta q^-_a ={1\over 4}\,(\epsilon^\beta_a D^-_\beta W +
\bar\epsilon^{\dot\alpha}_a\bar D^-_{\dot\alpha} \bar W)~,
\label{onshell} \eea where $\epsilon^{\alpha a}, \bar
\epsilon^{\dot\alpha a}$ are the Grassmann
transformation parameters. \\

\subsection{The Coulomb phase effective action}
We start with the calculation of ${\cal N}=4$
supersymmetric low-energy effective action extending the
non-holomorphic ${\cal N}=2$ superfield  potential (\ref{1}). This action
is assumed to have the following general form
\be
\Gamma[W,\bar{W},q^+]= \int d^{12}zdu
\left[{\cal H}(W,\bar{W})+
{\cal L}_q(W,\bar{W},q^+)\right]= \int d^{12}zdu\,
{\cal L}_{eff}
(W,\bar W, q^+)~. \label{9}
\ee
Here $d^{12} z$ is the full ${\cal N}=2$ superspace integration measure,
${\cal H}(W,\bar{W})$ is given by (\ref{1}), ${\cal L}_q (W, \bar W, q^+)$ is
some (for the moment unknown) function which should ensure, together with
${\cal H} (W, \bar W)$, the invariance of the functional \p{9} with respect to
the transformations \p{onshell}. Note that the Lagrangian
${\cal L}_q(W, \bar W, q^+)$, being a function of on-shell superfields,
must be in fact independent of the harmonics $u^\pm_i$.

The first term in \p{9} is transformed under \p{onshell} as
\be
\delta \int d^{12}zdu\, {\cal H} (W, \bar W) = \frac{1}{2} c \int
d^{12}zdu \frac{q^{+ a}}{\bar W W} ( \epsilon^\alpha_{a}
D^-_\alpha W + \bar \epsilon^{\dot \alpha}_{a} \bar D^-_{\dot
\alpha} \bar W )~. \label{var1}
\ee
Then ${\cal L}_q (W, \bar W, q^+)$ is to be determined from the condition that
its variation cancels the variation \p{var1}.

We introduce the quantity
\be
{\cal L}_{q}^{(1)} \equiv - c\,\frac{q^{+a}q^{-}_a}{\bar W W}
\ee
and notice that it transforms as
\be
\delta \,\frac{q^{+a}q^{-}_a}{\bar W W} = \frac{q^{+a}}{2\bar W W}
( \epsilon^\alpha_{\dot a} D^-_\alpha W +
\bar \epsilon^{\dot \alpha}_{\dot a} \bar D^-_{\dot \alpha} \bar W ) +
(q^{+a}q^{-}_a) \delta \left(\frac{1}{\bar W W}\right) +
D^{--}\left(\frac{ \delta
q^{+a}q^{+}_a}{\bar W W}\right). \label{10}
\ee
Then we consider
\be
{\cal L}_{eff}^{(1)} = {\cal H} (W, \bar W) - c\,\frac{q^{+a}q^{-}_a}{\bar W
W} =
  {\cal H} (W, \bar W) + {\cal L}_q^{(1)}~. \label{12}
\ee
Under the full harmonic ${\cal N} = 2$ superspace
integral the variation \p{var1} in ${\cal L}^{(1)}_{eff}$ is canceled by
the first term in \p{10}. The variation of \p{12} generated by the second
term in \p{10} remains non-canceled. After some algebra, it can be
brought into the form
\bea
\delta \int d^{12}zdu\,{\cal L}_{eff}^{(1)} &=& \frac{c}{2}
\int d^{12}zdu \frac{q^{+b}q^{-}_b}{(\bar W W)^2}
(\bar W  \bar \epsilon^{\dot \alpha a} \bar D^-_{\dot \alpha} q^+_a +
W \epsilon^{\alpha a} D^-_\alpha q^+_a)
\nn
&= & - \frac{c}{3} \int d^{12}zdu \frac{q^{+b}q^{-}_b}{(\bar W W)^2}q^{+a}
(\bar \epsilon^{\dot \alpha}_{a} \bar D^-_{\dot \alpha} \bar W   +
 \epsilon^{\alpha}_{a} D^-_\alpha W)~, \label{13}
\eea
where we have integrated by parts and used the off- and on-shell relations for $W, \bar W, q^\pm_a$ together with
cyclic identities for the SU(2) doublet indices.

Now let us consider the quantity
\be
{\cal L}_{eff}^{(2)} = {\cal L}_{eff}^{(1)} +
\frac{c}{3} \left(\frac{q^{+a}q^{-}_a}{\bar W W}\right)^2 \equiv
{\cal L}_{eff}^{(1)} +{\cal L}_{q}^{(2)}~,
\ee
where ${\cal L}_{eff}^{(1)}$ is given by \p{12}. The coefficient in the
new term  ${\cal L}_{q}^{(2)}$ has been picked up so that the variation
of the numerator of this term cancel \p{13}. The rest of the full
variation of ${\cal L}_q^{(2)}$ once again survives, and in
order to cancel it, one must add the appropriate  term ${\cal L}_{q}^{(3)}$
to ${\cal L}_q^{(1)} + {\cal L}_q^{(2)}$, and so forth.

The above consideration implies that the $q^{+a}$ dependent part of the
full effective action \p{9}, ${\cal L}_q = {\cal L}_q (W, \bar W,
q^{+})$, should be of the form
\be
{\cal L}_q = \sum^\infty_{n=1} {\cal L}_q^{(n)} = c
\sum^{\infty}_{n=1} c_n \left( \frac{q^{+a}q^{-}_a }{\bar W W} \right)^n
\label{17}
\ee
with some in advance unknown coefficients $c_n$. We have already specified
$c_1 = -1, c_2 = \frac{1}{3}$.
The further analysis proceeds by induction.

Let us consider two adjacent terms in the general expansion \p{17},
\be c_{n-1} \left( \frac{q^{+a}q^{-}_a }{\bar W W} \right)^{n-1} +
c_n  \left( \frac{q^{+a}q^{-}_a }{\bar W W} \right)^{n} \label{18}
\ee and assume that the variation of the numerator of the first term
has been already used to cancel the remaining part of the variation
of the preceding term [under the integral over the total  harmonic
superspace, as in \p{9}]. Then we rearrange the rest of the full
variation of the first term like in \p{13} and require this part to
be canceled by the variation of the numerator of the second term in
\p{18}. This results in the following recursive relation:
\be c_n = - 2
\frac{(n-1)^2}{n(n+1)} c_{n-1} \label{19}
\ee
 and $c_1 = -1$. This
immediately yields
\be c_n = \frac{(-2)^n}{n^2(n+1)}~.
\ee
As the
result, the full structure of  ${\cal L}_q$ is determined to be
\bea
{\cal L}_q (W, \bar W, q^+) &\equiv & {\cal L}_q (X) = c\,
\sum_{n=1}^{\infty} \frac{1}{n^2 (n+1)} X^n \nn &=&
c\left\{(X-1)\frac{\ln (1-X)}{X} + \left[ \mbox{Li}_2 (X) -1 \right]
\right\}~, \label{200}
\eea
where $X= -2\, \frac{q^{+a}q^{-}_a}{\bar W W}$ and $\mbox{Li}_2 (X)$ is the Euler dilogarithm. Let us
point out that the expression for $X$ does not depend on
harmonics due to the on-shell representation \p{onshq},
\be
X = -\frac{q^{ia}q_{ia}}{\bar W W}~.  \label{21}
\ee
Therefore, ${\cal L}_q(X)$ also does not depend on harmonics on shell and the integral
over harmonics in the effective action \p{9} can be omitted.

Thus, the full ${\cal N}=4$ supersymmetric low-energy effective action for
${\cal N}=4$ SYM model with gauge group SU(2) spontaneously broken down to
U(1) is given by
\bea
&& \Gamma [ W, \bar W, q^+] = \int d^{12}z\,
 {\cal L}_{eff} (W, \bar W, q^+)~, \label{23} \\
&& {\cal L}_{eff} (W, \bar W, q^+) = {\cal H} (W, \bar W) +
{\cal L}_q (X)\,,  \label{24}
\eea
where ${\cal H} (W, \bar W)$ and ${\cal L}_q (X)$ are given, respectively,
by \p{1} and \p{200}, and $X$ by \p{21} \footnote{
The functional \p{23} contains only quantum corrections. To write the whole
effective action, we have to add the classical action to the functional
\p{23}.}.

The expression \p{200} is the exact low-energy result. Indeed, the
non-holomorphic effective potential ${\cal H} (W, \bar W)$ \p{1} is exact,
as was argued in \cite{DS}. The Lagrangian ${\cal L}_q (X)$ \p{200} was
uniquely restored from \p{1} by ${\cal N}=4$ supersymmetry and it is the
only one forming, together with ${\cal H} (W, \bar W)$, an invariant of
${\cal N}=4$ supersymmetry. Therefore, the functional \p{23}, \p{24} is
the exact low-energy effective action for the theory under consideration.

Let us elaborate on the component structure
of the full effective action \p{23}, \p{24}. We consider only its bosonic
part, so that
\bea
& &W = \varphi(x) + 4 i \theta^+_{(\a} \theta^-_{\b )} F^{(\a\b)}(x)~, \quad
\bar W = \bar \varphi(x) +
4 i \bar \theta^+_{(\dot\a} \bar \theta^-_{\dot\b )}(x)
\bar F^{(\dot\a\dot\b)}(x)~, \quad q^{ia} = f^{ia}(x)~,
\nn
&& D^+_\a D^-_\b W = -4i F_{(\a\b)}~,
\quad\bar D^+_{\dot\a} \bar D^-_{\dot\b} \bar W =
4i \bar F_{(\dot\a\dot\b)}~.
\eea
Here $\varphi(x)$ is the complex scalar field of the ${\cal
N}=2$ gauge multiplet, $F^{\alpha\beta}(x)$ and $\bar
F^{\dot{\alpha}\dot{\beta}}(x)$ are the self-dual and anti-self-dual
components of the abelian field strength $F_{mn}$, and $f^{ia}(x)$
collects four scalar fields of the hypermultiplet $q^{ia}(z)$.
In this bosonic approximation, the functional argument $X$ \p{21} becomes
\be
X |_{\theta = 0} = - \frac{f^{ia}f_{ia}}{|\varphi|^2} \equiv X_0~.
\label{X0}
\ee
We ignore all $x$-derivatives of the involved
fields, since we are interested only in the leading part of
the expansion of the full effective action in the external momenta.

The component form of the effective action \p{23} can be straightforwardly
computed by performing integration over $\theta $s. After some computations we obtain
in bosonic sector the remarkably simple result
\be
\Gamma^{bos} = 4c \int d^4 x
\frac{F^2 \bar F^2}{(| \varphi|^2 + f^{ia}f_{ia})^2}~,
\label{boseff}
\ee
where $F^2 = F^{\alpha\beta}F_{\alpha\beta},
\bar F^2 = \bar F^{\dot{\alpha}\dot{\beta}}\bar
F_{\dot{\alpha}\dot{\beta}}$.
The expression in the denominator is none other
than the SU(4) invariant square of 6 scalar fields of ${\cal N} =4$
vector multiplet.  After proper
redefinitions, it can be cast in the manifestly SU(4) invariant form
\bea |\varphi |^2 + f^{ia}f_{ia} \sim \phi^{AB} \bar\phi_{AB}~,
\quad \phi^{AB} = - \phi^{BA}~, \quad \bar \phi_{AB} = \frac{1}{2}
\varepsilon_{ABCD} \phi^{CD}~, \quad A,B,C,D = 1,...,4~.  \nonumber
\eea This indicates that the effective action \p{23}, besides
being ${\cal N}=4$ supersymmetric, possesses also a hidden
invariance under the R-symmetry group SU(4)$_{\rm R}$ of ${\cal
N}=4$ supersymmetry.

The result \p{23} can be generalized for the theory with gauge group $\rm SU(N)$ spontaneously
broken down to $[\rm U(1)]^{N-1}$. In this case the effective action is given by the
general expression \p{200}, where ${\cal H}
(W, \bar W)$ has the form \p{2} and
\be
{\cal L}_{q} (W, \bar W, q^+) =
\sum_{I<J}{\cal L}_{q}^{IJ} (W, \bar W, q^+)~, \label{genLeff}
\ee
with each ${\cal L}_{q}^{IJ}$ being of the form \p{200}, in which $X$
should be replaced by
\bea
&& X_{IJ} = - 2 \frac{q^{+a}_{IJ} q^-_{aIJ}}{W_{IJ} \bar W_{IJ}} = -
\frac{q^{ia}_{IJ} q_{iaIJ}}{W_{IJ} \bar W_{IJ}}~,
 \label{genX} \\
&& W_{IJ} = W^I - W^J~, \quad \bar W_{IJ} =\bar W^I -\bar W^J~,
\quad q^{+a}_{IJ} = q^{+a}_I - q^{+a}_J~.
\label{defWq}
\eea
The hypermultiplet superfields are $q^{+a} = \sum_{I} q^{ +a}_{I}
e_{II}~,\; \sum_{I} q^{+a}_{I} =0~, $ and $e_{IJ}$ is the Weyl basis in
the algebra su(N). These hypermultiplet superfields belong to Cartan
subalgebra of su(N). In the SU(N) case the bosonic effective
action is represented by a sum of terms \p{boseff}.

As the final remark we note that the functional arguments $X$
\p{21}, \p{genX}  have the zero dilatation weight and are scalars of
the U(1) R-symmetry, since $q^{\pm a}$ and $W$ have the same dilatation
weights \cite{GIOS} and $q^{\pm a}$ behave as scalars under the R-symmetry
group. So, the full effective action \p{23} and its $su(N)$ analog  are expected to be invariant
under ${\cal N}=2$ superconformal symmetry like their pure $W, \bar W$ part
\p{1} or \p{2} \cite{BKT}. Being also ${\cal N}=4$ supersymmetric, these
actions respect the whole (on-shell) ${\cal N}=4$ superconformal
symmetry.

\setcounter{equation}{0}
\section{Low-energy effective action of $5D, \cN=2$ SYM theory}
In this section, we study the implications of extended supersymmetry
for the low-energy effective action of $5D$ SYM theory. This theory
is of interest from several points of view. It is non-renormalizable
by power-counting because of the dimensionful coupling constant $g$,
$[g]=-1/2$. Nevertheless, it was argued that a non-perturbative
quantum completion of this model describes $6D, \mathcal{N}=(2,0)$
superconformal field theory compactified on a circle
\cite{Douglas,LPS,LPS1}. An additional support to this conjecture
came from the exact computations of the partition function in this
theory by the localization technique (see e.g., the review
\cite{BUCH3} and the references therein).

In spite of the non-renormalizability of $5D, \mathcal{N}=1$ SYM, it
is still reasonable to study one-loop quantum corrections in it,
keeping in mind that in the odd-dimensional field theories
divergences can appear (within the dimensional regularization)
only at even loops. One-loop contributions to the effective action
of $5D, \mathcal{N}=1$ SYM theory were calculated in
refs.~\cite{Kuz06,Pletnev} for gauge group ${\rm SU}(2)$
spontaneously broken to ${\rm U}(1)$. The leading contribution is given
by the $5D$ supersymmetric Chern--Simons term \cite{Kuz06}, while
the next-to-leading one reads \cite{Pletnev}
\begin{equation}
\label{4deriv} c_{0}\int d^{5|8}z du \, W \ln \frac{W}{\Lambda }\,,
\end{equation}
where $W$ is the $5D, \mathcal{N}=1$ abelian gauge superfield
strength, $\Lambda $ is a scale parameter, $[\Lambda ]=1$, and the
integration is over the full $\mathcal{N}=1$ harmonic
superspace with the measure $d^{5|8}z du\equiv d^{5}x d^{8}\theta du$.
It is easy to check that the action (\ref{4deriv}) is $\Lambda
$-independent. The Chern--Simons term incorporates two-derivative
quantum corrections to the effective action, while (\ref{4deriv})
is $\mathcal{N}=1$ superfield extension of the four-derivative
``$F^{4}/\phi ^{3}$''-terms.

Our purpose is to study leading terms in the
low-energy effective action of $5D, \mathcal{N}=2$ SYM theory in
harmonic superspace. Although such terms might be found by direct
quantum computations in $5D, \mathcal{N}=1$ superspace, we
determine them here on the symmetry grounds, just by constructing
$\mathcal{N}=2$ completion of the $5D, \mathcal{N}=1$ SYM
effective action by the proper hypermultiplet terms.
The effective action constructed corresponds to the Coulomb branch
of $5D, \mathcal{N}=2$ SYM theory, with the gauge group being
broken to some abelian subgroup (e.g., the maximal torus), and, in
general, involves the massless abelian $\mathcal{N}=2$ gauge
multiplets valued in the algebra of this subgroup. For simplicity,
we focus on the case of the gauge group ${\rm SU}(2)$ and only
briefly address (in subsection 3.3) the case of ${\rm SU}(N)$ gauge
symmetry.

An additional motivation for studying the quantum effective action of  $5D, \cN=2$ SYM theory comes
from the D-brane stuff, like in the previous $4D, \cN=4$ example.
The classical action of $5D, \cN=1$ SYM theory with ${\rm U}(N)$ gauge group can
be interpreted as an action of a stack of $N$ D4-branes in flat
space-time. Then $\cN=2$ supersymmetric completion of the $5D, \cN=1$ SYM effective action
can be identified  with that of the four-derivative term in the low-energy effective action of
a single D4-brane on the $AdS_6\times S^4$ background.

\subsection{Classical action}
We start our consideration with a brief account of
the $\cN=1$ SYM and hypermultiplet models in $5D$ harmonic
superspace. We follow the notation and conventions of refs. \cite{KL,Pletnev}.

$\cN=2$ gauge multiplet in $5D, \cN=1$ harmonic superspace is
described by a pair of analytic superfields $(V^{++},q^+_a)$,
where $V^{++}$ is the $\cN=1$ gauge multiplet and
$q^+_a\equiv(q^+,-\bar q^+)$ is the hypermultiplet. The classical action of $V^{++}$ is written
as an integral over the full harmonic superspace
\cite{Zupnik87}
\be S_{\rm YM }= \frac1{2g^2} \sum_{n=2}^\infty
 \frac{(-i)^n}{n} \tr \int d^{5|8}z du_1 \ldots du_n
 \frac{V^{++}(z,u_1)V^{++}(z,u_2)\ldots V^{++}(z,u_n)}{(u^+_1 u^+_2)
 (u^+_2 u^+_3)\ldots
  (u^+_n u^+_1)}\,,
\label{Ssym}
\ee
$g$ being a coupling constant of dimension $-1/2$. The $V^{++}$ equation of motion reads
\be
({\cal D}^+)^2 W = 0\,,
\label{DW}
\ee
where $({\cal D}^+)^2 \equiv {\cal D}^{+\hat\alpha}{\cal D}^+_{\hat\alpha}$
and $W$ is a superfield strength of the gauge $\cN=1$ multiplet:
\be
W = \frac i8 ({\cal D}^+)^2 V^{--}\,.
\label{WV}
\ee
Here, the connection $V^{--}$ is related to $V^{++}$ through the
harmonic flatness condition
\be
D^{++} V^{--} - D^{--} V^{++} + i[V^{++},V^{--}] =0\,.
\label{zero}
\ee

The classical action of the hypermultiplet in the adjoint
representation of the gauge group is written as \cite{GIKOS,GIOS1,GIOS2}
\be
S_q = \frac1{2g^2} \tr \int d\zeta^{(-4)} q^+_a {\cal D}^{++}
q^{+a}\,,
\label{Sq}
\ee
where $d\zeta^{(-4)}$ is the  measure of integration over the analytic superspace and ${\cal D}^{++}
=D^{++}+i[V^{++}\,, \;$ is the gauge-covariant harmonic derivative. The equation of motion for $q^{+}_a$ is
\be
{\cal D}^{++} q^+_a =0\,.
\label{Dq}
\ee

The action of $\cN=2$ gauge multiplet in the $\cN=1$ harmonic
superspace formulation is just the sum of (\ref{Ssym}) and (\ref{Sq}),
\be
\label{S-class}
S^{\cN=2} = S_{\rm YM } + S_q\,.
\ee
This action is invariant under an implicit $\cN=1$ supersymmetry
\be
\delta q^+_a = -\frac12 (D^+)^4 [\epsilon_{a\hat\alpha}
 \theta^{-\hat\alpha} V^{--}]\,,\qquad
\delta V^{++} = \epsilon^a_{\hat\alpha} \theta^{+\hat\alpha}
q^+_a\,,
\label{hidden}
\ee
where $\epsilon^a_{\hat\alpha}$ is the relevant anticommuting parameter. Though the equation \p{DW} is modified for
the total action \p{S-class} by the hypermultiplet source term in the right-hand-side, it is not the case for the massless Cartan-subalgebra valued
abelian superfields which we will be interested in.
In the abelian case, the equations of motion for the $\cN=1$ gauge
multiplet (\ref{DW}) and hypermultiplet (\ref{Dq}) are simplified to the form
\be
(D^+)^2 W = 0\,,\qquad
D^{++} q^+_a = 0\,.
\label{EOM}
\ee
It is straightforward to show that on these equations the implicit
supersymmetry transformations (\ref{hidden}) are reduced to
\be
\delta q^\pm_a = \frac i2 \epsilon_a^{\hat\alpha}
(D^\pm_{\hat\alpha}
W)\,,\qquad
\delta W = -\frac i8 \epsilon^a_{\hat\alpha} D^{-\hat\alpha} q^+_a
+\frac i8 \epsilon^a_{\hat\alpha} D^{+\hat\alpha} q^-_a\,.
\label{hidden1}
\ee

\subsection{$\cN=2$ effective action}
In this subsection, we construct the complete low-energy effective action of
$\cN=2$ SYM theory with the gauge group ${\rm SU}(2)$ and both the gauge $\cN=1$ SYM and the hypermultiplet sectors taken into account.

The part of the superfield $\cN=1$ SYM effective action containing the component four-derivative term reads
\cite{Pletnev}
\be
\label{S011}
S_0 = c_0 \int d^{5|8} z du \, W\ln \frac W\Lambda\,,
\ee
where $W$ is the abelian gauge superfield strength, $\Lambda$ is a scale parameter, $c_0$ is a dimensionless constant and
the integration is performed over the
full $\cN=1$ harmonic superspace with the measure $d^{5|8}z du\equiv
d^5x d^8\theta du$. The representation (\ref{WV}) implies $\int d^{5|8} z du \, W = 0\,$, so the action (\ref{S011}) is
independent of the scale $\Lambda$, $dS_0/d\Lambda=0$.

The precise  value of the constant $c_0$ in the effective action (\ref{S011}) depends on the gauge group representation content of the hypermultiplet matter \cite{Pletnev}.
Here, we do not fix this constant and construct $\cN=2$ supersymmetric
generalization of (\ref{S0}), keeping $c_0$ arbitrary. This construction follows the same
steps as in ref. \cite{BuIv} (and in the previous section 2).

The variation of the action (\ref{S011})
under the hidden supersymmetry transformations
(\ref{hidden1}) may be cast in the form
\be
\delta S_0 = \frac{ic_0}{4} \int d^{5|8}z du \,
\epsilon^a_{\hat\alpha} q^+_a \frac{D^{-\hat\alpha}W }{W}\,.
\label{S0-var}
\ee
In deriving this equation we employed the abelian counterparts of
the relations (\ref{WV}), (\ref{zero}), the equations of motion
(\ref{EOM}) and integration by parts with respect to the harmonic and covariant spinor
derivatives.

The variation (\ref{S0-var}) may be partly canceled by the variation
of the action
\be
S_1 = c_1 \int d^{5|8}z du \frac{q^{+a}q^-_a}{W}\,,
\ee
where the coefficient $c_1$ will be defined below. The variation
of this action under (\ref{hidden1}) reads
\be
\delta S_1 = ic_1 \int d^{5|8}z du \frac{q^{+a}\epsilon^{\hat\alpha}_a(D^-_{\hat\alpha}W)}{W}
-\frac i8 c_1 \int d^{5|8}z du
 \frac{q^{+a}q^-_a}{W^2}(\epsilon^b_{\hat\alpha}D^{+\hat\alpha}q^-_b
  -\epsilon^b_{\hat\alpha}D^{-\hat\alpha}q^+_b)\,.
\label{deltaS1}
\ee
The first term in the right-hand side in (\ref{deltaS1}) cancels
the variation (\ref{S0-var}) if
\be
c_1 = -\frac{c_0}{4}\,,
\label{c1}
\ee
while the last term in (\ref{deltaS1}) may be cast in the form
\be
\delta (S_0 + S_1) = -\frac{i c_0}{12} \int d^{5|8}z du \frac{q^{+a}q^-_a}{W^3}
\epsilon^b_{\hat\alpha} q^+_b D^{-\hat\alpha}W\,.
\ee
To cancel this expression we are led to add the new term
\be
S_2 = c_2 \int d^{5|8}z du \frac{(q^{+a}q^-_a)^2}{W^3}\,,
\qquad
c_2 = \frac{c_0}{24}\,.
\label{S2}
\ee

Instead of evaluating the variation of the term (\ref{S2}) we
proceed to the general case and look for the full $\cN=2$ effective
action in the form
\be
S_{\rm eff}^{\cN=2}= \int d^{5|8}z du\left[
c_0 W \ln \frac W\Lambda +\sum_{n=1}^\infty c_n
\frac{(q^{+a}q^-_a)^n}{W^{2n-1}} \right]\,,
\label{Gamma}
\ee with
some coefficients $c_n$. Let us select two adjacent terms in the
sum in (\ref{Gamma}):
\be
c_n \frac{(q^{+a}q^-_a)^n}{W^{2n-1}} +
c_{n+1} \frac{(q^{+a}q^-_a)^{n+1}}{W^{2n+1}}\,.
\ee
It is possible
to show that the variation of the denominator in the first term
cancels the variation of the numerator in the second term, if the
coefficients are related as
\be (n+1) c_{n+1} = - c_n
\frac{n(2n-1)}{n+2}\,.
\ee
Taking into account Eq.\ (\ref{c1}), we
find from this recurrence relation the generic coefficient
\be c_n=
\frac{(-1)^n(2n-2)!}{n!(n+1)! 2^{n}}c_0\,.\label{cn}
\ee
This allows
us to sum up the series in (\ref{Gamma}) and to represent the
effective action in the closed form
\be
S_{\rm eff}^{\cN=2} = c_0
\int d^{5|8}z du\, W\left[ \ln \frac W\Lambda + \frac12 H(Z)
\right], \label{Seff}
\ee
where
\be
Z\equiv \frac{q^{+a} q^-_a}{W^2}
\,,
\ee
and
\be
H(Z) = 1+ 2\ln\frac{1 + \sqrt{1+2Z}}{2} +\frac23
\frac{1}{1 + \sqrt{1+2Z}} -\frac43\sqrt{1+2Z}\,. \label{H}
\ee
It is
easy to check that $H(0) =0\,, \;H^\prime(0) = -\frac12$, in
agreement with \p{cn}. The result (\ref{Seff}) was derived in the work \cite{5DBIS}.

The action (\ref{Seff}) is $\cN=2$ supersymmetric
extension of the effective action (\ref{S011}). It would be
interesting to reproduce this result from the perturbative quantum
computations in $5D$ harmonic superspace, like it has been done in the $4D$ case
in \cite{BIP,BBP,BP}.

It is worth to point out that the term (\ref{4deriv}) we have started with (as well as
its analogs for the higher-rank gauge groups) may arise in quantum
theory only as a one-loop quantum correction to the effective
action. Indeed, it is scale-invariant and so is independent of the
gauge coupling constant $g$. On the other hand, within the
background field method in harmonic superspace \cite{BBKO,Pletnev},
all higher-loop Feynman graphs involve a gauge superfield vertex
with the coupling constant $g$. Thus, all higher-loop quantum
contributions to the effective action are not scale-invariant and
for this reason cannot give rise to a renormalization of the
coefficient $c_0$ in Eq.\ (\ref{4deriv}). However, in contrast to
the $4D$ case, this coefficient is not protected against
non-perturbative corrections. Such corrections will be discussed
elsewhere.

It is straightforward to generalize this
result to a  higher-rank gauge group. For instance, for the ${\rm SU}(N)$ gauge group
spontaneously broken to the maximal torus
$[{\rm U}(1)]^{N-1}$ we obtain
\be
S_{\rm eff}^{\cN=2} = c_0 \sum_{I<J}^N \int d^{5|8}z du\, W_{IJ}\left[
\ln \frac{W_{IJ}}{\Lambda} + \frac12 H(Z_{IJ})
\right], \label{HSUn}
\ee
where $Z_{IJ}= \frac{(q^{+a})_{IJ} (q^-_a)_{IJ}}{W^2_{IJ}}$ and
$W_{IJ}=W_I - W_J$, $(q^{\pm a})_{IJ}=q^{\pm a}_{I}-q^{\pm
a}_{J}$. The superfields $W_I$ and $q^{\pm a}_{I}$ obey the constrains
$\sum_I W_I = 0$, $\sum_I q^{\pm a}_{I} =0$ and span the
Cartan directions in the algebra ${\rm su}(N)$. The function $H(Z_{IJ})$
for each argument $Z_{IJ}$ is given by the expression \p{H}.

\subsection{Component structure}
We will be interested in deriving the term $F^4/\phi^3$ from the
effective action (\ref{Seff}). To this end, it is enough
to leave only the following component fields in the involved superfields:
\bea
q^+&=&f^i(x)u^+_i\,,\quad
\bar q^+=-\bar f^i(x) u^+_i\,,\\
W&=&\sqrt2\phi(x) - 2i
\theta^{+\hat\alpha}\theta^{-\hat\beta}F_{\hat\alpha\hat\beta}(x)\,.
\label{Wcomp}
\eea
Here $\bar\phi = \phi$, $\overline{f^i} = \bar f_i$ are scalar
fields and $F_{\hat\alpha\hat\beta} = F_{\hat\beta\hat\alpha}$ is
Maxwell field strength.

Substituting (\ref{Wcomp})
into the first term in (\ref{Seff}), we find
\be
S_0 = c_0\frac{\sqrt2}{3}\int d^{5|8}z \frac{(\theta^{+\hat\alpha}\theta^{-\hat\beta}F_{\hat\alpha\hat\beta})^4}{\phi^3}
=\frac{c_0}{4\sqrt2}\int d^{5|8}z\frac{\det
F}{\phi^3}(\theta^+)^2(\theta^+)^2(\theta^-)^2(\theta^-)^2\,,
\label{S01}
\ee
where $\det F = \frac1{4!}\varepsilon^{\hat\alpha\hat\beta\hat\gamma\hat\delta}
\varepsilon^{\hat\mu\hat\nu\hat\rho\hat\sigma}F_{\hat\alpha\hat\mu}
F_{\hat\beta\hat\nu}F_{\hat\gamma\hat\rho}F_{\hat\delta\hat\sigma}$
and $(\theta^\pm)^2 =
\theta^{\pm\hat\alpha}\theta^\pm_{\hat\alpha}$. We integrate over
the Grassmann variables according to the rule
\be
\int d^{5|8}z \,(\theta^+)^2(\theta^+)^2(\theta^-)^2(\theta^-)^2
f(x) = 4\int d^5x\,f(x)\,,
\ee
for some $f(x)$. Thus the action (\ref{S01}) yields the
component term
\be
S_0 = \frac{c_0}{\sqrt2} \int d^5x \frac{\det F}{\phi^3}\,.
\label{S02}
\ee

In a similar way one can find the contribution of the last term in (\ref{Seff})
\be
\int d^{5|8}z \,W H(Z) =
\sqrt2\int d^5x \frac{\det F}{\phi^3}
[4 z^4 H^{(4)}(z)+28 z^3 H'''(z)+ 39 z^2 H''(z)+6z H'(z)]\,,
\label{2.29}
\ee
where
\be
z\equiv Z|_{\theta=0} = \frac{f^i \bar f_i}{\phi^2}\,.
\ee
Substituting the function (\ref{H}) into (\ref{2.29}), we find
\be
\frac{c_0}{2} \int d^{5|8}z \,W H(Z) =
\frac{c_0}{\sqrt2} \int d^5x \frac{\det F}{(\phi^2 + f^i\bar f_i)^{3/2}}
-\frac{c_0}{\sqrt2} \int d^5x \frac{\det F}{\phi^3}\,.
\ee
The last term exactly cancels (\ref{S02}). As a
result, the total $F^4/\phi^3$ term in the component form of the
effective action (\ref{Seff}) is given by the expression
\be
S_{\rm eff}^{\cN=2} =\frac{c_0}{\sqrt2} \int d^5x \frac{\det F}{(\phi^2 +f^i \bar
f_i)^{3/2}}+\ldots,
\label{Scomp}
\ee
where dots stand for the remaining terms. It is remarkable that the scalar fields
appear in the denominator in (\ref{Scomp}) just in the ${\rm SO}(5)$
invariant combination. This is non-trivial, since the field $\phi$ comes
from the gauge $\cN=1$ multiplet, while $f^i, \bar f_i$ from the
hypermultiplet. In the ${\rm SU}(N)$ case \p{HSUn}, $S^{{\cal N}=2}_{\rm eff}$ is a sum of the appropriate terms \p{Scomp}.

\setcounter{equation}{0}
\section{Low-energy effective action of $6D, \cN=(1,1)$ SYM theory}
Another interesting class of extended objects in superstring/brane
theory is presented by D5-branes (see e.g., \cite{GK,BKLS}). These
objects are related to $6D,\cN=(1,1)$ SYM theory likewise  D3-branes
are related to $4D,\,\cN=4$ SYM theory. Similarly to the D3-brane
case, the interaction of D5-branes is described by $6D$ Born-Infeld
action \cite{T} (see e.g., \cite{Sevrin2,DHHK,GTKK} and the reference therein  for aspects of the Born-Infeld
action in diverse dimensions). Since D5-brane is related to
$6D,\cN=(1,1)$ SYM theory, it is natural to expect that the D5-brane
interaction in the low-energy limit can be calculated proceeding
from the low-energy quantum effective action of this theory.

In this section we consider quantum aspects of
$6D,\,\cN=(1,1)$ SYM theory. It is the maximally extended
supersymmetric gauge theory in six dimensions, with eight
left-handed and eight right-handed supercharges. An equal number of
spinors with mutually-opposite chiralities guarantees  the absence
of chiral anomaly in this theory. From the point of view of $6D,\,
\cN=(1,0)$ supersymmetry, the model is built on a gauge (vector)
multiplet and a hypermultiplet. Respectively, its bosonic sector involves
a real $5D$ gauge field and two complex (or
four real) scalar fields.

Although $6D,\, \cN=(1,1)$ non-abelian SYM theory is
non-renormalizable by power counting, it is
on-shell finite at one and two loops
\cite{FT,MarSag1,MarSag2,HS,HS1,BHS,BHS1,Bork}. Moreover, it was
recently shown that this theory is one-loop finite even off-shell
\cite{BIMS-a,BIMS-b,BIMS-c} and that the two-loop diagrams with
hypermultiplet legs are also off-shell finite \cite{BIMS-d}. Review
of our approach was presented in \cite{BIMS-review}.

To preserve as many manifest supersymmetries as possible we use the
harmonic superspace approach \cite{GIKOS, GIOS}. The theory under
consideration is formulated in terms of $\cN=(1,0)$ harmonic
superfields describing the gauge multiplet and the hypermultiplet.
Therefore it possesses the manifest $\cN=(1,0)$ supersymmetry and,
in addition, a non-manifest (hidden) $\cN=(0,1)$
supersymmetry mixing $\cN=(1,0)$ gauge multiplet and hypermultiplet.
These supersymmetries close on shell on the total on-shell $\cN=(1,1)$
supersymmetry. Such a formulation of $\cN=(1,1)$ SYM theory was
described in detail in the paper \cite{BIS} (see also ref.
\cite{HSW,Z}). An essential difference of our consideration here is the use
of the so called ``$\omega$-form'' of the hypermultiplet (see
below).

We consider the case when gauge symmetry ${\rm SU}(N)$
is broken to ${\rm SU}(N-1)\times {\rm U}(1) \subset {\rm SU}(N)$. Technically, this
means that background superfields align through the fixed generator
of Cartan subalgebra of ${\rm SU}(N)$,  which corresponds to an abelian
subgroup ${\rm U}(1)$. In this case the effective action of the theory
depends only on the abelian vector multiplet and hypermultiplet. In the
bosonic sector we find out the effective action for the single real ${\rm U}(1)$
gauge field and four real scalar fields. The same number of bosonic
world-volume degrees of freedom is exhibited by a single
D5-brane in six dimensions \cite{S_98}.

\subsection{$6D,\, \cN=(1,1)$ SYM in the ${\cal N}=(1,0)$ harmonic formulation with
the $\omega$ hypermultiplet}
We start with the formulation of $6D, \,\cN=(1,1)$ SYM theory in terms
of $6D,\, \cN=(1,0)$ harmonic analytic superfields $V^{++}$ and
$\omega$, which represent  the gauge multiplet and the
hypermultiplet. The action of $\cN = (1,1)$ SYM theory is written
as
 \bea
S_0[V^{++}, q^+]&=&
\frac{1}{\rm f^2}\Big\{\sum\limits^{\infty}_{n=2} \frac{(-i)^{n}}{n} \tr \int
d^{14}z\, du_1\ldots du_n \frac{V^{++}(z,u_1 ) \ldots
V^{++}(z,u_n ) }{(u^+_1 u^+_2)\ldots (u^+_n u^+_1 )}  \nn
&& - \frac12 \tr \int d\zeta^{(-4)}\, \nb^{++}\omega \nb^{++}\omega
\Big\}\,,\label{S0}
 \eea
where ${\rm f}$ is a dimensionful coupling constant ($[{\rm f}]=-1$)
and the measure of integration over the analytic subspace
$d\zeta^{(-4)}$ includes the integration over harmonics,
$d\zeta^{(-4)} = d^6 x_{(\rm an)}\, du\,(D^-)^4$. Both $V^{++}$ and
$\omega$ superfields take values in the adjoint representation of
the gauge group. The covariant harmonic derivative $\nb^{++}$ acts
on the hypermultiplet $\omega$ as
 \be
\nabla^{++}\omega = D^{++} \omega + i [V^{++},\omega]\,. \label{Vfirst}
 \ee
The action \p{S0} is invariant under the infinitesimal gauge transformations
 \bea
 \d V^{++} = -\nb^{++} \Lambda\,,  \quad \d \omega =  i[\Lambda, \omega]\,,
 \label{gtr}
 \eea
where $\Lambda(\zeta, u) = \widetilde{\Lambda}(\zeta, u)$ is a real
analytic gauge parameter.

Besides the analytic gauge connection $V^{++}$ we introduce a
non-analytic one $V^{--}$  which is a solution of the zero curvature
condition \cite{GIOS}
 \bea
 D^{++} V^{--} - D^{--}V^{++} + i [V^{++},V^{--}]=0\,. \label{zeroc}
 \eea
Using $V^{--}$, we define one more covariant harmonic
derivative $\nb^{--} = D^{--} + i V^{--}$ and the ${\cal
N}=(1,0)$  gauge superfield strength
 \bea
 W^{+a}=-\frac{i}{6}\varepsilon^{abcd}D^+_b D^+_c D^+_d V^{--}
 \eea
possessing the useful off-shell properties
 \bea
\nabla^{++} W^{+a}  = \nabla^{--} W^{-a} \ =\ 0\,, \qquad W^{-a} =
\nabla^{--}W^{+a} \,. \label{HarmW}
 \eea

Introducing an analytic superfield $F^{++}\,$,
 \bea
 F^{++} = \frac14 D^{+}_a W^{+ a} = i(D^+)^4 V^{--}\,, \qquad D^+_a F^{++} = \nb^{++} F^{++}=0\,,
 \eea
we can write the classical equations of motion corresponding to the action \p{S0} as
 \bea
 F^{++}  +[\omega,\nabla^{++}\omega] = 0\,, \qquad  (\nabla^{++})^2\,\omega = 0\,.
\label{Eqm}
 \eea

The $\cN=(1,0)$ superfield action \p{S0} enjoys the additional $\cN=(0,1)$ supersymmetry
 \bea
\delta V^{++} &=& (\epsilon^{+ A}u^+_A) \omega - (\epsilon^{+
A}u^-_A) \nabla^{++}\omega
= 2 (\epsilon^{+ A}u^+_A) \omega - \nabla^{++} \big((\epsilon^{+ A}u^-_A)
 \omega\big),\label{V++HidOm} \\
\delta \omega &=& -(D^+)^4 \big((\epsilon^{-A}u^-_A) V^{--}\big) =
i(\epsilon^{- A}u^-_A) F^{++} -i (\epsilon^A_a u^-_A)\, W^{+ a}
\label{omegaHid},
 \eea
where $A=1,2$ is the  Pauli-G\"{u}rsey ${\rm SU}(2)$ index. To check this,
one derives,  using
\p{V++HidOm} and \p{omegaHid},  ${\cal N}=(0,1)$ transformation law of
$\nb^{++}\omega\,$,
 \bea
\delta(\nabla^{++}\omega) = i\big( (\epsilon^{-A}u^+_A)  +
(\epsilon^{+A}u^-_A)\big)F^{++} -i (\epsilon^A_a u^+_A)\, W^{+ a}
+ i(\epsilon^{+ A}u^-_A)[\omega, \nabla^{++}\omega]. \label{omega3Hid}
 \eea
Then  one varies the classical action \p{S0} with respect to \p{V++HidOm} and
\p{omega3Hid}
 \bea
 \d S = \frac{1}{\rm f^2}  \Big\{\tr\int d^{14} z du \,V^{--} \d V^{++} -
 \tr\int d \zeta^{(-4)} \nb^{++} \omega\, \d( \nb^{++}
 \omega)\Big\}\,.
 \eea
In the first integral, we pass to the integration over the analytic subspace and use
the explicit form of the variations  \p{V++HidOm} and \p{omega3Hid}
 \bea
\d S &=& -\frac{i}{\rm f^2}  \tr\int d \zeta^{(-4)} \Big\{ 2F^{++}
(\epsilon^{+ A}u^+_A) \omega +\nb^{++}\omega \big(
(\epsilon^{-A}u^+_A)+(\epsilon^{+A}u^-_A)\big)F^{++} \nn
 && -F^{++}\nabla^{++}
\big((\epsilon^{+ A}u^-_A) \omega \big) - \epsilon^A_a u^+_A\,
\nb^{++}\omega\, W^{+ a} \Big\} =0\,. \label{var}
 \eea
The last two terms in \p{var} are the total harmonic derivative
$\nb^{++}$ due to the properties of $F^{++}$ and $W^{+a}$ and so they vanish
under the analytic integration measure $ d \zeta^{(-4)}$. The first two terms cancel each other
after integration by parts with respect to the harmonic derivative $\nb^{++}$  and
using the properties $\nb^{++} \epsilon^{-A} = \epsilon^{-A}$ and $\nb^{++} u^{-}_A
= u^{+}_A$. Finally, the term  $\tr \big( \nb^{++}\omega
[\omega, \nb^{++}\omega]\big)$ vanishes due to the cyclic property of trace.

The zero curvature condition \p{zeroc} allows one to express the
transformation of the non-analytic gauge connection $\delta V^{--}$
through $\d V^{++}$
 \be
 \nabla^{++}\delta V^{--} - \nabla^{--}\delta V^{++} = 0\,,
 \ee
and to find the transformation low of the strength
$W^{+a}$ under the hidden supersymmetry
 \bea
 \delta W^{+ a} =\varepsilon^{adbc} \epsilon^A_d \nabla_{bc}
  \big(u^+_A \omega - u^-_A \nabla^{++}\omega\big)
 + i\epsilon^{- A}[W^{+ a}, u^+_A \omega - u^-_A \nabla^{++}\omega],
 \label{WHid}
 \eea
where
 \bea
\nabla_{bc} = \partial_{bc} -\frac12 D^+_b D^+_c
\,V^{--}\,.
 \eea

As usual, we make use of the background superfield method. The gauge group of the theory \p{S0} is assumed to be ${\rm SU}(N)$. For the further
consideration, we will also assume that the background superfields ${\bf
V}^{++}$ and  ${\bf \Omega}$ align in a fixed direction in the
Cartan subalgebra of ${\rm su}(N)$
 \bea
 {\bf V}^{++} = V^{++}(\zeta,u) H\,, \qquad {\bf \Omega} = \Omega(\zeta,u)\, H\,,
 \eea
where $H$ ia a fixed generator in  the Cartan subalgebra generating
some abelian subgroup ${\rm U}(1)$. Our choice of the background corresponds to
the spontaneous symmetry breaking ${\rm SU}(N) \rightarrow {\rm SU}(N-1)\times
{\rm U}(1)$. We have to note that the pair of the background superfields
$(V^{++},\Omega)$ forms an abelian vector $\cN=(1,1)$ multiplet
which, in the bosonic sector, contains  a single real gauge vector
field $A_M(x)$ and four real scalars $\phi(x)$ and
$\phi^{(ij)}(x)\,, i,j=1,2$, where $\phi$ and $\phi^{(ij)}$ are the
scalar components of $\Omega$ hypermultiplet \cite{GIOS}. The
abelian vector field and four scalars in six-dimensional space-time
constitute just the bosonic world-volume degrees of freedom of a
single D5-brane \cite{GK,BKLS}.

The classical equations of motions \p{Eqm} for the background
superfields $V^{++}$ and  $\Omega$ are:
 \bea
 F^{++} = 0\,, \qquad (D^{++})^2 \Omega = 0\,. \label{EqmB}
 \eea
In that follows we assume that the background superfields solve the
classical equation of motion \p{EqmB}. We will also consider the
background slowly varying in space-time, {\it i.e.},
 \bea
 \partial_M W^{+a} = 0\,, \qquad \partial_M \Omega = 0\,.
 \label{constBG}
 \eea

Finally, we are left with an abelian background analytic superfields  $V^{++}$
and $\Omega$, which satisfy the classical equation of motion
\p{EqmB} and the conditions \p{constBG}. Under these assertions the
gauge superfield strength $W^{+a}$ is analytic \footnote{ In
general this is not true and $F^{++}\neq0$.}, $D^+_a W^{+b} =
\d_a^b F^{++}=0$. In further analysis we will use the
$\cN=(0,1)$ transformation for gauge superfield strength $W^{+a}$
\p{WHid}. For the slowly varying abelian on-shell
background superfields the hidden $\cN=(0,1)$ supersymmetry
transformations \p{omegaHid} and \p{WHid} acquire the very simple form
 \bea
  \delta \Omega = -i (\epsilon^A_a u^-_A)\, W^{+ a}\,, \qquad \delta W^{+ a}=0.
  \label{Onshell2}
 \eea
These transformation rules follow from the abelian version of the
transformations  \p{omegaHid}, \p{WHid} in which one has to take
into account the conditions  \p{constBG} and \p{EqmB}. It is worth
pointing out that these conditions are covariant under
${\cal N}=(0,1)$ supersymmetry.

\subsection{Effective action with hidden ${\cal N}=(0,1)$ supersymmetry}
Let us now consider the simplest $\cN=(1,1)$
invariants which can be constructed out of  the abelian analytic superfields
$W^{+a}$ and $\Omega$ under the assumptions \p{EqmB} and
\p{constBG}. It is evident that the following gauge-invariant
action
 \bea
I={\rm f}^2 \int d\zeta^{(-4)} (W^{+})^4 {\cal F}({\rm f}\Omega),
\label{I}
 \eea
where $(W^+)^4 =
-\frac{1}{24}\varepsilon_{abcd} W^{+a}W^{+b}W^{+c}W^{+d}$ and ${\cal
F}({\rm f}\Omega)$ is an arbitrary function of $\Omega$,
is invariant under the transformation \p{Onshell2} due to the
nilpotency condition $(W^{+})^5\equiv0$. For our further consideration, of the main interest
is the choice
 \bea
I_1= c\int d\zeta^{(-4)}\, \frac{(W^{+})^4}{ \Omega^2}\,, \label{I1}
 \eea
which corresponds to ${\cal F} = \frac{1}{{\rm f^2}\Omega^2}$ in
\p{I}. The coefficient $c$ in \p{I1} cannot be fixed only on the
symmetry grounds and should be calculated in the framework of the
quantum field theory \cite{Buchbinder:2017xjb}. In fact, the same concerns the specific choice of the function
${\cal F}({\rm f}\Omega)$.

As a result of the exact calculations \cite{Buchbinder:2017xjb}, we arrived at the following expression for the one-loop effective action:
 \bea
\Gamma^{(1)}_{\rm lead} = \frac{N-1}{(4\pi)^3}\int d \zeta^{(-4)}\,
\frac{(W^+)^4}{\Omega^2}\,. \label{1loop4}
 \eea
As expected, the leading low-energy contribution \p{1loop4}  to the
effective action in the model \p{S0} is just the $\cN=(1,1)$
invariant $I_1$ \p{I1}. The coefficient $c$ now takes the
value \cite{Buchbinder:2017xjb}
 \bea
 c=\frac{N-1}{(4\pi)^3}\,. \label{c}
 \eea
The expression fore $c$ is similar to that in the four-dimensional $\cN=4$ SYM theory
(see, e.g., \cite{KU} and references therein). In the bosonic sector
the effective action \p{1loop4} has the structure
 \bea
\Gamma^{(1)}_{\rm bos} \sim \int d^6 x\, \frac{F^4}{\phi^2}\bigg(1+
\frac{\phi^{(ij)}\phi_{(ij)}}{\phi^2}+\ldots\bigg)\,,
 \eea
where $F^4 = F_{MN} F^{MN} F_{PQ} F^{PQ} - 4 F^{NM} F_{MR} F^{RS}
F_{SN}$ and $F_{MN}$ is the gauge field strength.

We have to note that even in the one-loop approximation there might
exist more complicated contributions to the effective action, which
are beyond the scope of our consideration. We hope to come back to these issues elsewhere.

\setcounter{equation}{0}
\section{Summary and outlook}
In this paper we have reviewed a hidden-supersymmetry based  approach to constructing the
low-energy effective actions for extended and higher-dimensional
supersymmetric gauge theories. We considered the $4D, \cN=4;\,5D,
\cN=2;$ and $6D, \cN=(1,1)$ SYM theories formulated in harmonic
superspace. All these theories are characterized by some number of
manifest off-shell supersymmetries and some number of hidden
on-shell supersymmetries. The complete supersymmetry of the theories
under consideration is due to a combination of the manifest and hidden
ones. The low-energy effective actions were analyzed on the purely algebraic grounds and can
be obtained in a general form up to numerical coefficients. To fix
these coefficients, one needs to carry out the concrete quantum field-theoretical calculations.
This approach has been completely accomplished for all three theories under consideration.

In $4D, \cN=4$, ${\rm SU}(N)$ SYM theory we begun with the known
manifestly $\cN=2$ supersymmetric non-holomorphic effective potential in the gauge multiplet sector
and, using the hidden supersymmetry transformations, completed it by hypermultiplet terms to
the full $\cN=4$ supersymmetric low-energy effective potential in Coulomb
phase \cite{BuIv}. The result was later confirmed by the one-loop
supergraph calculations \cite{BIP,BBP,BP}.

Generalizing the approach of ref. \cite{BuIv} to $5D$ case, we
constructed the leading term in the low-energy effective action of
$5D, \cN=2$ SYM theory as the appropriate sum of the effective
action of  $5D, \cN=1$ SYM theory and the interactions with the
hypermultiplet. This interaction is fixed, up to an overall coupling
constant $c_0$, by the requirement of the implicit on-shell $5D,
{\cal N}=1$ supersymmetry extending the manifest off-shell ${\cal
N}=1$ supersymmetry to an on-shell $5D, {\cal N}=2$ one. We
discussed in detail the case of the gauge group ${\rm SU}(2)$
spontaneously broken to ${\rm U}(1)$, in which case the effective action
depends on a single pair of abelian $5D, {\cal N}=1$ gauge multiplet
and hypermultiplet, and then considered a more general situation
with the ${\rm SU}(N)$ gauge group broken to its maximal torus, with $N-1$
pairs of such abelian multiplets \cite{5DBIS}.

The next obvious problem is to reproduce the $5D$ effective
actions found in \cite{5DBIS} from the appropriate set of quantum $5D, {\cal N}=1$
supergraphs involving the interacting hypermultiplet and ${\cal
N}=1$ gauge superfields. Also, it would be interesting to establish
manifest links with the relevant D-brane dynamics and the $4D$ and
$6D$ cousins of the $5D$ effective action constructed. Finding out
an explicit form of the next-to-leading corrections to this
effective action, also based  on the demand of implicit $5D, {\cal
N}=1$ supersymmetry, is another interesting task.

In $6D, \cN=(1,1)$ SYM theory we dealt with  its formulation in terms of $\cN=(1,0)$ harmonic
superfields as the theory of interacting
$\cN=(1,0)$ gauge multiplet and hypermultiplet, both in the adjoint representation
of the gauge group. This theory is characterized by
manifest $\cN=(1,0)$ supersymmetry and hidden on-shell $\cN=(0,1)$ supersymmetry.
The low-energy effective action has been constructed by combining the gauge multiplet
superfield strengths and hypermultiplet, so as to achieve an invariance under
the hidden supersymmetry. The result has been actually confirmed by a direct quantum computation \cite{Buchbinder:2017xjb}.

The results concerning the low-energy effective actions and the roles of hidden supersymmetries
in the above three theories can be generalized along many directions. First,
in all cases we found only the leading low-energy effective actions, omitting all the superspace
derivative-dependent terms. However, such terms could in principle be essential for establishing the precise links
with superstring/brane low-energy effective actions. Second, it would be interesting to sum up all
superfield strength-dependent terms without derivatives and to obtain in this way the Heisenberg-Euler or Born-Infeld
type effective actions possessing both manifest and hidden supersymmetries. The third interesting
direction is related to exploring the quantum structure of the higher-derivative
supersymmetric theories, e.g., of $6D$ renormalizable
higher-derivative supersymmetric gauge theory \cite{IvSmZu, TC}.

\section*{Acknowledgements} This work was supported by the RFBR grant No. 18-02-01046.
E.A.I. thanks the Editors of the Andrei Slavnov Jubilee Volume for invitation to submit the contribution. This review paper is partly
based on joint works with Boris Merzlikin, Albert Petrov, Nikolai Pletnev, Igor
Samsonov and Konstantin Stepanyantz. We are  sincerely indebted to them.


\end{document}